\documentstyle[aps,preprint]{revtex}
\input{epsf}
\begin{document}
\draft
\title{ A Variational Principle for Eigenvalue Problems of
 Hamiltonian Systems}
\author{R.\ D.\ Benguria and M.\ C.\ Depassier}
\address{        Facultad de F\'\i sica\\
	P. Universidad Cat\'olica de Chile\\ Casilla 306,
Santiago 22, Chile}

\maketitle
\begin{abstract}
We consider the bifurcation problem $u'' + \lambda u = N(u)$ with two
point boundary conditions  where $N(u)$ is a general  nonlinear term
which may also depend on the eigenvalue $\lambda$.  We give a
variational characterization of the bifurcating branch $\lambda$ as a
function of the amplitude of the solution. As an application we show
how it can be used to obtain simple approximate closed formulae for
the period of large amplitude oscillations.
\end{abstract}

\pacs{2.30.Hq, 3.20+i, 2.30.Wd}

Different physical systems reduce to the consideration of a one
dimensional equation of the form $ u'' = F(u,\mu)$, where the force
$F$ depends on $u$ and one or more parameters $\mu$. In these systems
the energy $E = {1\over 2}{u'}^2 + V(u)$ is a constant of the motion
and the qualitative dynamics can be understood from the analysis of
the phase plane and the stability of the fixed points. Among the
systems that fall in this class we may cite the buckling of a thin
rod, the pendulum, steady state solutions of the reaction diffusion
equation and numerous others \cite{Hale2,Keller69,Nayfeh,Rabin77}
(Given the large number of references in this area we refer to some
of the standard textbooks only).  We shall assume that the system has
an equilibrium point at $u = 0$. For small deviations from the
equilibrium point the behavior of the solution is determined by the
linear equation $u'' + \lambda u = 0$ with suitable boundary
conditions.   An important feature of the linear problem is that the
eigenvalue $\lambda$ does not depend on the amplitude of the motion.
For large deviations from the equilibrium, the full problem, which we
write as
\begin{equation}
u'' + \lambda u = N(u,\lambda)
\label{eq:primera}
\end{equation}
 must be considered. A solution to this equation and the analytic
determination of the eigenvalue $\lambda$ is possible only for
special forms of the nonlinearity $N$. When an analytic solution is
not possible, one can resort to obtaining a numerical solution, or
use perturbation theory around the linear problem.  The eigenvalues
of the linear part of Eq.(\ref{eq:primera}) correspond to the points
where bifurcation occurs in the nonlinear problem.  There exist
different methods of perturbation theory applicable to these systems
\cite{Minorsky,Nayfeh}. An important property of the solution in the
nonlinear problem is that the eigenvalue $\lambda$ depends on the
amplitude of the solution.

The purpose of this article is to show that the eigenvalue
derives from a variational principle. The main tool used in the
derivation of this result is a new variational characterization
of the linear problem which, for the sake of clarity, we
consider first. 

We wish to solve the linear problem
\begin{mathletters}
 \begin{equation}
u'' + \lambda u = 0 \label{eq:linear}
\end{equation}
subject to
\begin{equation}
 u'(0) = 0, \qquad u(1) = 0
\end{equation}
\end{mathletters}
where the boundary conditions we have chosen are the ones 
appropriate for a
wide range of problems. We refer to other choices at the end.
We will denote $u(0) = u_m$. The positive 
solution to this problem is given by $u_m \cos ( \pi x/2)$
corresponding to  the lowest eigenvalue $\lambda = (\pi/2)^2$.
 There exists a
variational principle for this equation, the Rayleigh-Ritz
principle which provides upper bounds on $\lambda$. This
principle however cannot be extended to calculate the eigenvalue
of nonlinear problems. We shall construct a new one whose
extension to nonlinear problems will be straightforward.

Let $g(u)$ be an arbitrary positive function such that $g(0) =
0$ and $g'(u) > 0$.
Multiplying Eq.(\ref{eq:linear}) by $ u'g(u)$ and integrating in
$x$ we obtain, after integrating by parts,
\[
-{1\over 2} \int_0^1 (u')^2 g'( u) u' dx + \lambda \int_0^1 u
g(u) u' \, dx = 0
\]
where the boundary terms vanish since $u'(0) = 0$ and $ g[u(1)]=
g(0) = 0$. From the equation and the boundary conditions it
follows that $u'< 0$ in $(0,1)$ so in the second integral we may
use $u$ as the independent variable. We obtain then the
identity, valid for any admisible $g$,
\begin{equation}
\lambda \int_0^{u_m} u
g(u) du = - {1\over 2} \int_0^1 (u')^2 g'( u) u' dx.
\label{eq:identidad}
\end{equation}

Consider now, for a fixed $g$ the functional
\begin{mathletters}
\begin{equation}
J_g[v] =  - {1\over 2} \int_0^1 (v')^2 g'( v) v' dx \equiv
{1\over 2} \int_0^1 \Phi(v,v') dx
\label{eq:funcional}
\end{equation}
defined for functions $v$ which satisfy
\begin{equation}
v(0) = u_m, \qquad v(1) = 0, \quad \mbox{and}  \quad v' < 0 
 \quad \mbox{in} \quad
(0,1). 
\label{eq:vborde}
\end{equation}
\end{mathletters}
Then $J_g[v] \ge 0$ 
 and for fixed $g$ it has a unique minimum attained at $v =
\tilde v$. A rigorous proof will be given elsewhere. 
The minimizing function $\tilde v$ satisfies the Euler-Lagrange
equation for $J_g$. Since $\Phi (v,v')$ does not depend on $x$ the
Euler-Lagrange equations can be integrated once to yield $\Phi - v'
\partial\Phi / \partial v' = \mbox{constant} $. Since $\Phi(v,v') = -
{1\over 2} v^3 g'(v)$ we get
\begin{equation}
(\tilde v')^3 g'(\tilde v) = - K \label{eq:vtilde}.
\end{equation}
Then, given $g$ we may determine $\tilde v$ by integrating
Eq.(\ref{eq:vtilde}) subject to the boundary conditions (\ref{eq:vborde}).
We have then, for fixed $g$,
\begin{equation}
J_g[v] \ge \min J_g[v] = J_g[\tilde v] = {1\over 2} K(u_m) \label{eq:cotaJ}
\end{equation}
where the dependence of $K$ on $u_m$ is obtained through the
boundary conditions on $\tilde v$.

Before going any further, consider for example $g(v) = v$. Then,
the minimizing $v$,  $\tilde v$, satisfies $\tilde v'(x) = -
K^{1/3}$. Imposing   the boundary
conditions we get $\tilde v = K^{1/3} (1 - x)$ provided that $K =
u_m^3$. For this choice of $g$, therefore, $J_g \ge u_m^3/2$.

Now we go back to the problem under consideration. We have from
equations (\ref{eq:identidad}), (\ref{eq:funcional}) and
(\ref{eq:cotaJ}),
\begin{eqnarray}
\lambda \int_0^{u_m} u
g(u) du &=& - {1\over 2} \int_0^1 (u')^2 g'( u) u' dx \nonumber \\
        &\ge& J_g[\tilde v ] = - {1\over 2} \int_0^1 (\tilde v')^2
           g'( \tilde v)
\tilde v' dx  \label{eq:desigual} \\
       &=& {1\over 2} K(u_m) \nonumber
\end{eqnarray}
which gives an upper bound on $\lambda$.  We have shown that for
 any given $g(u)$ (with $g(0)= 0,\, g'(u)>0$),
\begin{equation}
\lambda \ge {1\over 2} { K(u_m) \over \int_0^{u_m} u g(u)\, du }.
\label{eq:bound}
\end{equation}
When will equality hold in Eq.(\ref{eq:bound})? From
Eq.(\ref{eq:desigual}) we see that the equality will hold when $g$ is
chosen in such a way that  $\tilde v$ coincides with $u$. For this
$g$ which we will call $\hat g$ we will have
\begin{equation}
u'^3 \hat g'(u) = - K. \label{eq:ghat}
\end{equation}
Since $u$ is a solution of the differential equation (\ref{eq:linear})
we know
that 
\begin{equation}
 {1\over 2} u'^2 + {1\over 2} \lambda u^2 = E = {1\over 2}\lambda
u_m^2  \label{eq:energy}
\end{equation}
from where it follows  $ {u'}^2 =
\lambda u_m^2 ( 1 - (u/u_m)^2)$. Replacing this in 
Eq.(\ref{eq:ghat}) we obtain
\[
\hat g'(u) = {K\over \left[ (\lambda u_m^2) (1 - u^2/u_m^2)\right]^{3/2}}.
\]
This equation can be integrated to yield (omitting overall
multiplicative constants)
\begin{equation}
\hat g (u) = {u \over (1 - (u/u_m)^2 )^{1/2} }.
\end{equation}
Our final result for the linear problem is then the following
variational characterization for the lowest eigenvalue,
\begin{equation}
\lambda = \max {1\over 2} { K(u_m) \over \int_0^{u_m} u g(u)\, du },
\end{equation}
where the maximum is taken over all positive functions $g$ such that
$g(0) = 0$, $g'> 0$. The maximum is attained when $g =  \hat g $.

It is straightforward to evaluate the integrals for $g = \hat g$
and verify that we obtain the correct result. Solving $ {v'}^3 \hat
g' = - K$ with $v(0) = u_m$, $v(1) = 0$ we obtain $K = (\pi
/2)^3 u_m^3$ and $\int u g(u) du = (1/2) (\pi /2) u_m^3$ and the
exact  value for $\lambda$ is obtained.

We now consider the nonlinear problem
\begin{mathletters}
 \begin{equation}
u'' + \lambda u = N(u) \label{eq:entera}
\end{equation}
subject to
\begin{equation}
 u'(0) = 0, \qquad u(1) = 0
\end{equation}
\end{mathletters}
As before we denote $u(0) = u_m$.
Again multiplying by $u'\,g(u)$, where $g$ is as before, we obtain,
after integrating
\begin{equation}
\lambda \int_0^{u_m} u
g(u) du =  \int_0^{u_m} N(u) g(u) du 
- {1\over 2} \int_0^1 (u')^2 g'( u) u' dx.
\end{equation} 
where in the term involving $\lambda$ and the nonlinearity we have
used $u$ as the independent variable. The second term in the right
side is just the linear term considered above so we have
\begin{equation}
\lambda \ge {1\over 2} {\int_0^{u_m} N(u) g(u) du 
+  K(u_m) \over \int_0^{u_m} u g(u)\, du }
\end{equation}
and, as before the maximum is attained when $g$ is chosen in such a
way that $ \tilde v = u$. Now the equation for $\hat g$ becomes, 
\begin{equation}
\hat g'(u) = {1\over (E - V(u))^{3/2} }
\label{eq:ge}
\end{equation}
where $V(u)$ is the potential. Here the potential $V = {1\over 2}
\lambda u^2 - \int_0^u N(y)\, dy$ and $E ={1\over 2} \lambda u_m^2 -
\int_0^{u_m} N(y)\, dy$  This expression for $\hat g$ cannot be
integrated in general due to the nonlinear terms in $V(u)$ that arise
from $N(u)$. However, as in the linear case, the maximizing $g$
exists and we obtain our main result
\begin{equation}
\lambda = \max {1\over 2} {\int_0^{u_m} N(u) g(u) du 
     +  K(u_m) \over \int_0^{u_m} u g(u)\, du }
\label{eq:main}
\end{equation}
where the maximum is taken over all positive functions $g$ such that
$g(0) = 0$ and $g'(u) > 0$. The maximum is attained for $g = \hat g$.

Since in general we will not be able to integrate Eq.(\ref{eq:ge})
the above principle gives lower bounds which can be arbitrarily close
to the exact value.  

As an example we shall use as a trial function the function $g$ which
gives the correct eigenvalue for the linear problem $ g(u) = u /
\sqrt{1 - (u/u_m)^2}$. With this trial function we obtain from
(\ref{eq:main}) the lower bound
\begin{equation}
\lambda \ge \left({\pi \over 2}\right)^2 + {4\over \pi u_m^3} \int_0^
{u_m} { N(u) u \over \sqrt{1 - u^2/u_m^2} } du
\label{eq:formula}
\end{equation}
which is valid for any $N(u)$. 

As a first example consider the equation
$$
u''+ \lambda u = u^2 - {7\over 5} u^3 + {1\over 2} u^4
$$
with $u(0) = u_m$, $u'(0) = 0$, $u(1) = 0$. The integral in
Eq.(\ref{eq:formula}) can be done easily. We obtain
$$
\lambda \ge \left( {\pi \over 2}\right)^2 + {4\over \pi}\left( 
{2\over 3} u_m - {3\over 8}\cdot {\pi\over 2}\cdot {7\over 5}u_m^2
 + {1\over 2}\cdot {8\over 15} u_m^3 \right).
$$
In Fig. 1 we show a plot of $u_m(\lambda)$. The solid line
corresponds to the exact value obtained from the numerical
integration of the equation. The dashed line corresponds to the lower
bound given by the formula above. As it can be seen in the figure,
the lower bound gives a very accurate value for amplitudes up to $u_m
\approx 2$. For larger values there is a small discrepancy which can be
minimized by a more adequate choice for $g$. The formula given in any
case is a close enough lower bound even at larger amplitudes.
\vbox{
\epsfxsize=13truecm \epsfysize=8.5truecm
\epsfbox{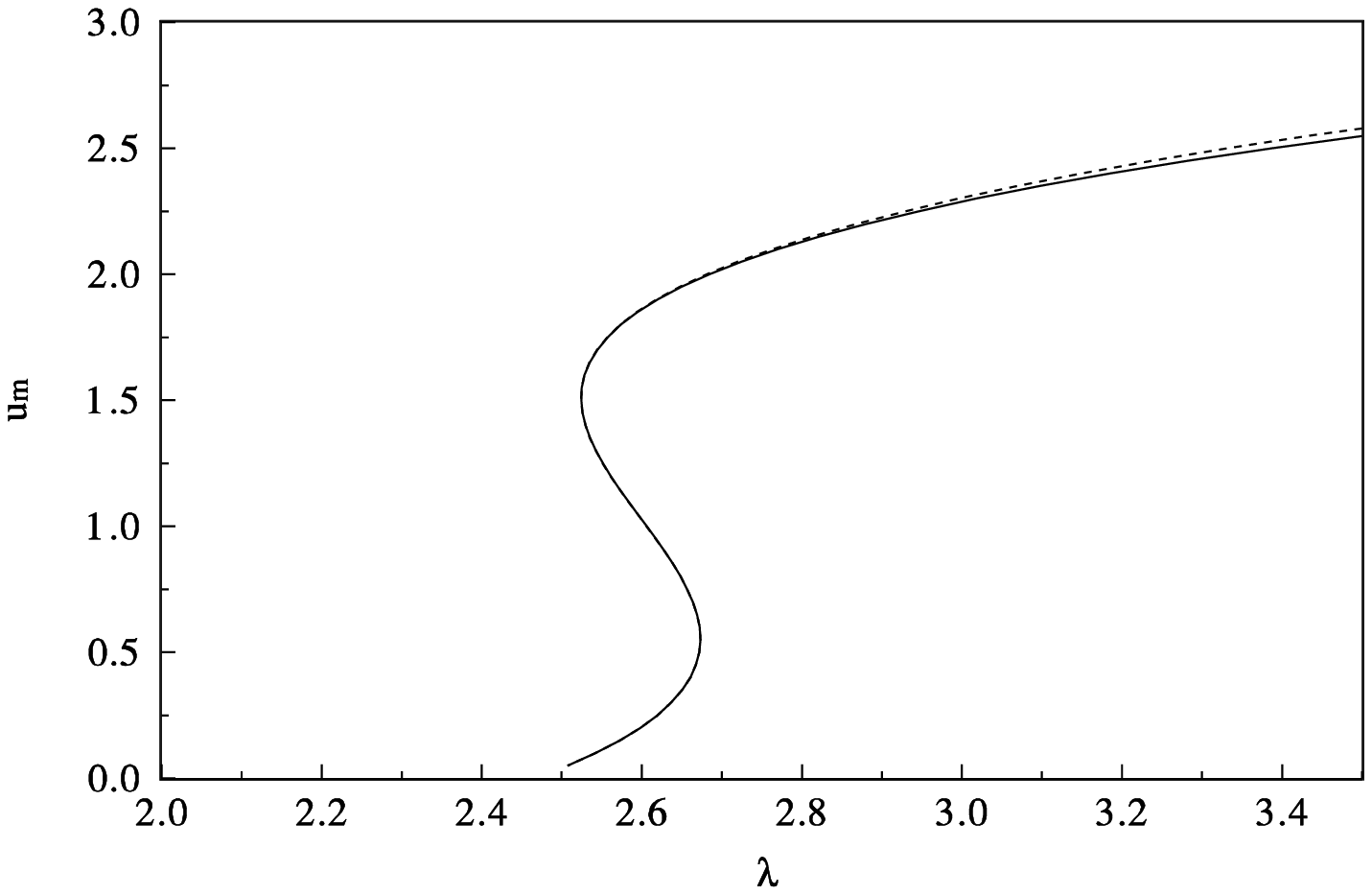}
}
\begin{centerline}{ Fig. 1} \end{centerline}

As a second example we consider the unforced Duffing equation
$$
\ddot x + x + \delta x^3 = 0.
$$
For this equation $x= 0$ is a stable equilibrium point and we search
for the period of oscillation of perturbations from the equilibrium.
This is an exactly solvable equation and the period can be given in
terms of elliptic functions. When the nonlinearity is small,
perturbation theory can be used to calculate the period or frequency
of oscillations. The frequency obtained assuming $\delta \ll 1$ is 
$$
\omega_{pert} = {2\pi \over T} = 1 + {3\over 8} \delta a^2 
- {15\over 256} \delta^2 a^4 + \mbox{higher order terms}
$$
where $a$ is the amplitude of the oscillations. In order to make use
of the variational result we recall that since the potential has the
symmetry $x \rightarrow - x$ we may consider a quarter period of the
solution which we chose as the lower right quadrant in phase space
$(\dot x, x)$. In this region of phase space we have $\dot x(0) = 0$,
$x(0) = a$, and $x(T/4) = 0$.  Then, introducing the scaled time
variable $\tau = 2 \omega t/\pi $ we obtain 
$$
 {d^2 x\over d \tau^2}
+ \left({\pi\over 2\omega}\right)^2 (x + \delta x^3) = 0, 
$$
 with
 $$
x(0) = a, \qquad \dot x(0) = 0, \qquad x(1) = 0.  
$$
 We identify
$\lambda = (\pi / 2\omega )^2$, $N(x) = - \delta \lambda x^3$ and
apply  Eq.(\ref{eq:formula}). We obtain the bound
 $$
\lambda \ge \left( {\pi\over 2}\right)^2 - {3\over 4} \delta \lambda a^2,
$$
which written in terms of $\omega$ gives,
\begin{equation}
\omega_{var} \le \sqrt{ 1 + {3\over 4}\delta a^2}\, .
\end{equation}
This bound is valid for oscillations of any amplitude. In Fig. 2 we
show the results for $\delta = 0.1$. The solid line gives the exact
solution, the dashed line the variational bound and the dot-dashed
line the perturbation result including terms up to order $\delta^2$.
Again we see the close agreement up to fairly large amplitude of the
variational expression. As in the previous example a different trial
function would give a close estimate at larger amplitudes.
\vbox{
\epsfxsize=13truecm \epsfysize=8.5truecm
\epsfbox{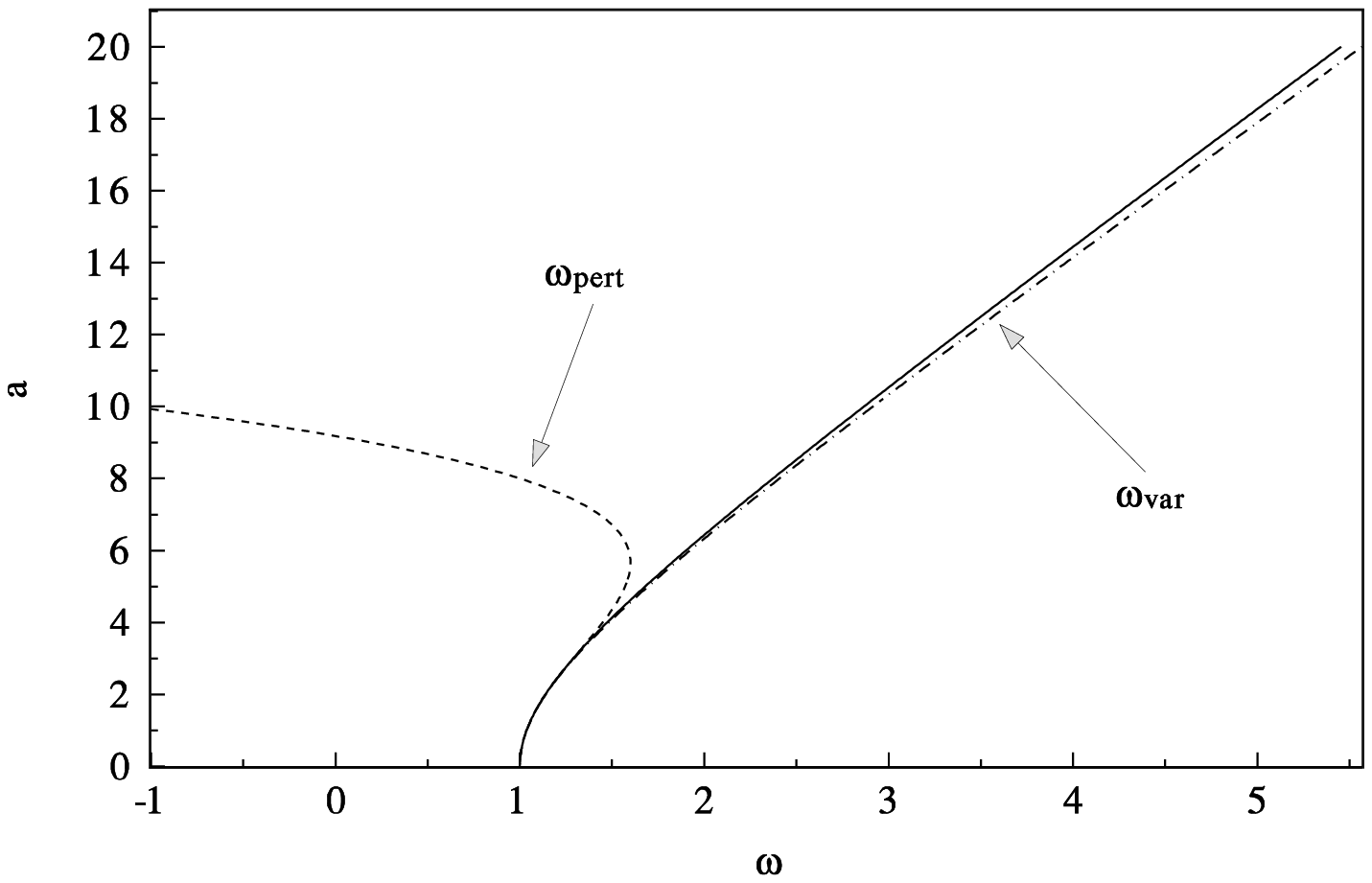}
}
\begin{centerline}{ Fig. 2} \end{centerline}

In this last example we were able to reduce to a quarter period of
the solution due to the symmetry of the potential. In the absence of
such symmetry we should consider half the period, that is the lower
half of the phase plane. The relevant bifurcation problem to be
considered in this case is 
$$
 u'' + \lambda u = N(u) 
$$
 subject to
 $$
u(0) = u_m, \qquad u'(0) = 0, \qquad u'(1) = 0. 
 $$
 One can obtain a
similar variational principle in this case too. The details are
slightly more involved but entirely similar to what we have shown.

To conclude, we have shown that nonlinear eigenvalue problems for
hamiltonian systems derive from a variational principle. The main
tool in the derivation of this result is a new variational
formulation of the linear problem. The variational principle may be
used to calculate the eigenvalues as accurately as desired by
suitable choice of trial functions. A simple estimate, using as a
trial function the optimal trial function for the linear problem,
gives reasonably accurate results for small to medium amplitude
solutions, and it gives a good  rigorous bound for larger amplitudes.
The method that we have used here can be applied to other problems.
In particular it can be shown to be a  reformulation in real space of
the method used previously by us \cite{BDCMP,preprint} to obtain the
speed of fronts of a reaction--diffusion equation in one dimension
for arbitrary nonlinearities.  Due to the nature of the front problem
though, its formulation directly in phase space is simpler. It is an
open question whether this approach can be extended to treat limit
cycles of nonlinear oscillators.

This work has been partially supported by Fondecyt project 1960450. 
R.B. was supported by a C\'atedra Presidencial.

\end{document}